\documentclass[11pt,twocolumn,tight,times]{aastex62}
\usepackage{graphicx,color}
\usepackage{mathrsfs,amsmath,amssymb}
\usepackage{ulem}
\usepackage{url}
\def\bhm{M_{\bullet}}

\def\feii{Fe {\sc ii}}

\def\hb{H$\beta$}

\def\kms{\rm km~s^{-1}}

\def\oiii{[O\,{\sc iii}]}

\def\sunm{M_\odot}

\newcommand{\broad}{\ifmmode {\xi} \else $\xi$ \fi}

\def\heii{He~{\sc ii}}

\def\calA{{\cal A}}

\def\calW{{\cal W}_\varPsi}
\def\calWW{{{\cal W}_C}}

\defcitealias{du2014}{Paper~I} 
\defcitealias{wang2014a}{Paper~II}
\defcitealias{hu2015}{Paper~III} 
\defcitealias{du2015}{Paper~IV} 
\defcitealias{du2016a}{Paper~V}   
\defcitealias{du2016b}{Paper~VI} 

\begin{document}

\title{A HIGH-QUALITY VELOCITY-DELAY MAP OF THE BROAD-LINE REGION IN NGC~5548}
	
\correspondingauthor{Pu Du, Jin-Ming Bai}
\email{dupu@ihep.ac.cn, baijinming@ynao.ac.cn}

\author{Ming Xiao}
\affiliation{Yunnan Observatory, Chinese Academy of Sciences, Kunming 650011, Yunnan, China}
\affiliation{Key Laboratory for Particle Astrophysics, Institute of High Energy Physics,
	Chinese Academy of Sciences, 19B Yuquan Road, Beijing 100049, China}
\affiliation{University of Chinese Academy of Sciences, 19A Yuquan Road, Beijing 100049, China}

\author{Pu Du}
\affiliation{Key Laboratory for Particle Astrophysics, Institute of High Energy Physics,
	Chinese Academy of Sciences, 19B Yuquan Road, Beijing 100049, China}

\author{Kai-King Lu}
\affiliation{Yunnan Observatory, Chinese Academy of Sciences, Kunming 650011, Yunnan, China}

\author{Chen Hu}
\affiliation{Key Laboratory for Particle Astrophysics, Institute of High Energy Physics,
	Chinese Academy of Sciences, 19B Yuquan Road, Beijing 100049, China}

\author{Yan-Rong Li}
\affiliation{Key Laboratory for Particle Astrophysics, Institute of High Energy Physics,
	Chinese Academy of Sciences, 19B Yuquan Road, Beijing 100049, China}

\author{Zhi-Xiang Zhang}
\affiliation{Key Laboratory for Particle Astrophysics, Institute of High Energy Physics,
	Chinese Academy of Sciences, 19B Yuquan Road, Beijing 100049, China}

\author{Kai Wang}
\affiliation{Key Laboratory for Particle Astrophysics, Institute of High Energy Physics,
	Chinese Academy of Sciences, 19B Yuquan Road, Beijing 100049, China}

\author{Ying-Ke Huang}
\affiliation{Key Laboratory for Particle Astrophysics, Institute of High Energy Physics,
	Chinese Academy of Sciences, 19B Yuquan Road, Beijing 100049, China}

\author{Jin-Ming Bai}
\affiliation{Yunnan Observatory, Chinese Academy of Sciences, Kunming 650011, Yunnan, China}

\author{Wei-Hao Bian}
\affiliation{Physics Department, Nanjing Normal University, Nanjing 210097, China}

\author{Luis C. Ho}
\affiliation{Kavli Institute for Astronomy and Astrophysics, Peking University, Beijing 100871, China} 
\affiliation{Department of Astronomy, School of Physics, Peking University, Beijing 100871, China} 

\author{Ye-Fei Yuan}
\affiliation{Department of Astronomy, University of Science and Technology of China, Hefei 
	230026, China}

\author{Jian-Min Wang}
\affiliation{Key Laboratory for Particle Astrophysics, Institute of High Energy Physics,
	Chinese Academy of Sciences, 19B Yuquan Road, Beijing 100049, China}
\affiliation{University of Chinese Academy of Sciences, 19A Yuquan Road, Beijing 100049, China}
\affiliation{National Astronomical Observatories of China, Chinese Academy of Sciences,
	20A Datun Road, Beijing 100020, China}

\received{2018 August 3}
\revised{2018 September 3}
\accepted{2018 September 5}
\journalinfo{To appear in {\it The Astrophysical Journal Letters}.}
	
\begin{abstract}
NGC 5548 has been well spectroscopically monitored for reverberation mapping of the central 
kinematics by 19 campaigns. Using the maximum entropy method in this Letter, we build up a 
high-quality velocity-delay map of the \hb\ emission line in the light curves of the continuum 
and the line variations observed between 2015-2016. The map shows the response strength and 
lags of the velocity fields of the \hb\ emitting regions. The velocity-delay structure of the 
map is generally symmetric, with strong red and blue wings at time lag $\tau \lesssim 15$ days, 
a narrower velocity distribution at $\tau \gtrsim 15$ days, and a deficit of response in the 
core. This is suggestive of a disk geometry of the broad-line region (BLR).
The relatively weaker \hb\ response at the longer lags in the red side indicates anisotropic 
emission from the outer part of the BLR. We also recover the velocity-delay maps of NGC~5548 
from the historical data of 13 years to investigate the long-term variability of its BLR. In 
general, the BLR of NGC 5548 was switching between the inflow and virialized phases in the 
past years. The resultant maps of seven years reveal inflow signatures and show decreasing 
lags, indicating that the changes in the BLR size are related to the infalling BLR gas. 
The other four maps show potential disk signatures which are similar to our map.
\end{abstract}
	

\section{Introduction}
Broad emission lines are the most prominent features of type 1 active galactic nuclei (AGNs). 
It is now generally accepted that these Doppler-broadened lines arise from the clouds in the 
broad-line region (BLR) photoionized by the central continuum radiation. Reverberation mapping 
(RM, see \citealt{bahcall1972,blandford1982,peterson1993}) has been widely used to investigate 
the geometry and kinematics of the BLRs in AGNs 
(\citealt{kaspi2000,peterson2004,denney2010,bentz2010a,grier2012,du2018}). The variations of the 
continuum ($\Delta C$) and the emission line ($\Delta L$) are connected as the expression:
\begin{equation}
\Delta L(v,t)=\int_{0}^{\infty}\varPsi(v,\tau)\Delta C(t-\tau)d\tau,
\label{RMmodel}
\end{equation}
where $\varPsi(v,\tau)$ is the so-called ``transfer function" or velocity-delay map. The map gives 
the distribution of the line response over the line-of-sight velocity $v$ and the time delay $\tau$,
and can be used to indicate the nature of the BLR.

Due to the limitation of data quality, early attempts mainly focus on analyzing the time lags as a 
function of the line-of-sight velocity (velocity-resolved time-lag analysis, see, e.g., 
\citealt{barth2011a,barth2011b,bentz2008,bentz2009,bentz2010a,denney2009a,denney2009b,denney2010,
	doroshenko2012,du2016,grier2013,lu2016,	pei2017,rosa2015,ulrich1996}). 
Velocity-resolved time-lag analysis has been applied to more than two dozen sources, and 
preliminarily reveals their BLR geometry and kinematics. However, this method measures the mean time 
lags in each velocity bins rather than recover the velocity-delay maps which can reveal the detailed
response features. Subsequently, more advanced techniques such as the maximum entropy method (MEM, 
\citealt{horne1991,horne1994,xiao2018}), regularized linear inversion method (RLI, 
\citealt{krolik1995,skielboe2015}) and the dynamical modeling method 
(\citealt{pancoast2011,pancoast2014a,li2013}) begin to be applied to recover the velocity-delay maps
(\citealt{bentz2010b,grier2013,grier2017, pancoast2012,pancoast2014b,pancoast2018,ulrich1996,xiao2018}).
Briefly speaking, MEM and RLI directly recover the velocity-delay map without adopting any specific 
model for the geometry and dynamics of the BLR.
Dynamical modeling method fits the variations of the emission line profile by assuming specific BLR 
model, which can provide the velocity-delay map and the black hole mass measurement simultaneously.
 
As one of the most intensively observed objects in the RM study, NGC~5548 has been monitored by 19 
individual campaigns, including the International AGN Watch Consortium (\citealt{peterson2002}, and 
references therein), \cite{bentz2007}, \cite{denney2009b}, the 2008 Lick AGN Monitoring Project 
(LAMP2008; \citealt{bentz2009}), the AGN Space Telescope and Optical Reverberation Mapping 
(AGN STORM; \citealt{rosa2015,edelson2015,fausnaugh2016,pei2017}) and \cite{lu2016}.
Three velocity-resolved time-lag analyses for the \hb~line of NGC~5548 were presented, which reveal a
velocity-symmetric line response  and suggest that the BLR of NGC~5548 tends to be virialized 
or a Keplerian disk \citep{denney2009b,lu2016,pei2017}. 
The velocity-resolved time-lag measurement of LAMP2008 data shows similar time lags in each velocity 
bins \citep{bentz2009}, but the corresponding dynamical modeling analysis suggests that the dynamics 
of the BLR is dominated by inflow \citep{pancoast2014b}. The velocity-delay map of LAMP2008
recovered by the RLI analysis is in general consistent with the velocity-resolved result, but
shows a prompt response in the red wing \citep{skielboe2015}.
In addition, \cite{kollatschny2013} investigated the BLR geometry through the profiles of the 
emission lines of NGC 5548, and proposed that its BLR conforms to the disk wind model
(\citealt{murray1997,proga2004}), and the geometry tends to be not thick.

In 2015, we started an RM campaign of NGC~5548 to investigate its BLR physics. 
The mean and velocity-resolved time lags have been presented in \cite{lu2016}. This Letter 
recovers the velocity-delay map of NGC~5548 by using MEM. In addition, to study the long-term 
variations of its BLR kinematics and geometry, we also analyze the AGN watch data of 13 
years and recover the corresponding velocity-delay maps.

\section{observations and data reduction}
\label{secob}
The spectroscopic data used in this Letter were obtained in our RM campaign during 2015 January-July, 
by using the Yunnan Faint Object Spectrograph and Camera mounted in the 2.4 m telescope at Lijiang 
Station of Yunnan Observatories of Chinese Academy of Sciences.
The details of the observation and data reduction were provided in \cite{lu2016}. 
For completeness, we briefly introduce some general points here. (1) We oriented the 
long slit (2\arcsec.5 wide) to take the spectra of NGC~5548 and a
nearby comparison star simultaneously, and calibrate the spectra of the object by using the 
comparison star. This procedure gives a calibration accuracy of $\sim$ 2\% for NGC~5548
(see more details in \citealt{lu2016}); (2) The instrument broadening 
is $\sim500\ \kms$ (see \citealt{du2016}), which is much less than the FWHM of the \hb~line in the 
mean spectrum ($\sim9912\pm362\ \kms$, see Table 4 in \citealt{lu2016}); (3) The time span of this 
observation is 205 days (with 61 epochs); (4) We used a spectral fitting scheme to measure the 
continuum and \hb~line fluxes (see details in \citealt{hu2015,lu2016}), which allows us to remove 
the irrelevant spectral components as the \heii~and \feii~emission, the narrow lines, and the host 
galaxy. In the MEM analysis of our observation, we use the broad \hb\ profiles after subtracting 
those irrelevant components.

\begin{figure}
	\centering
	\includegraphics[angle=0,width=0.41\textwidth]{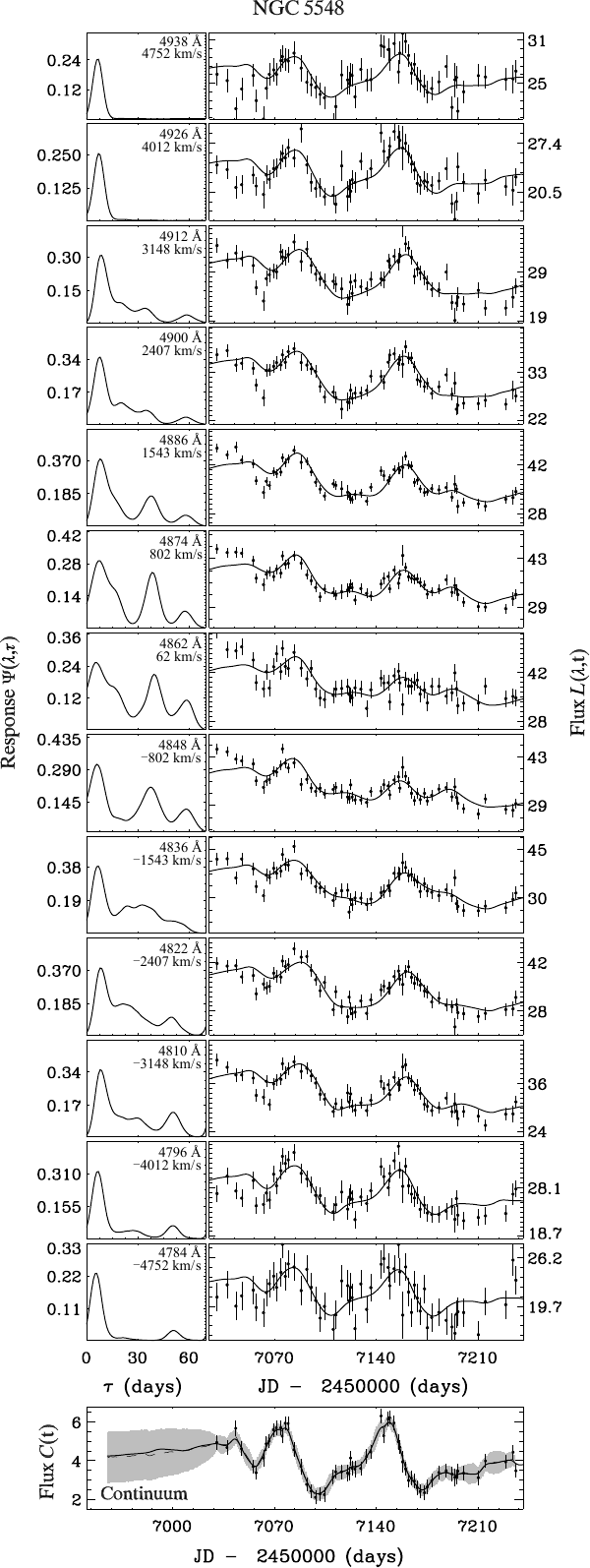} 
	\caption{\footnotesize
		Some examples of the light curve fitting at different wavelengths (in the rest frame) 
		for NGC 5548. The left panels show the corresponding 1-dimensional response functions. 
		The bottom panel shows the original continuum light curve (black dots with error bars), 
		the MEM reconstruction of the continuum (solid black line), and a damped random walk 
		modeling (dashed line with gray envelope
		) for comparison. The selected 	wavelengths are labeled along the top axis of Figure 
		\ref{map}. $C(t)$ and $L(\lambda, t)$ are in units of 
		$10^{-15}\ {\rm erg\ s^{-1}\ cm^{-2}\ \AA^{-1}}$ and 
		$10^{-16}\ {\rm erg\ s^{-1}\ cm^{-2}\ \AA^{-1}}$, 
		respectively.
	}
	\label{spec}
\end{figure}

\begin{figure}
	\centering
	\includegraphics[angle=0,width=0.3\textwidth]{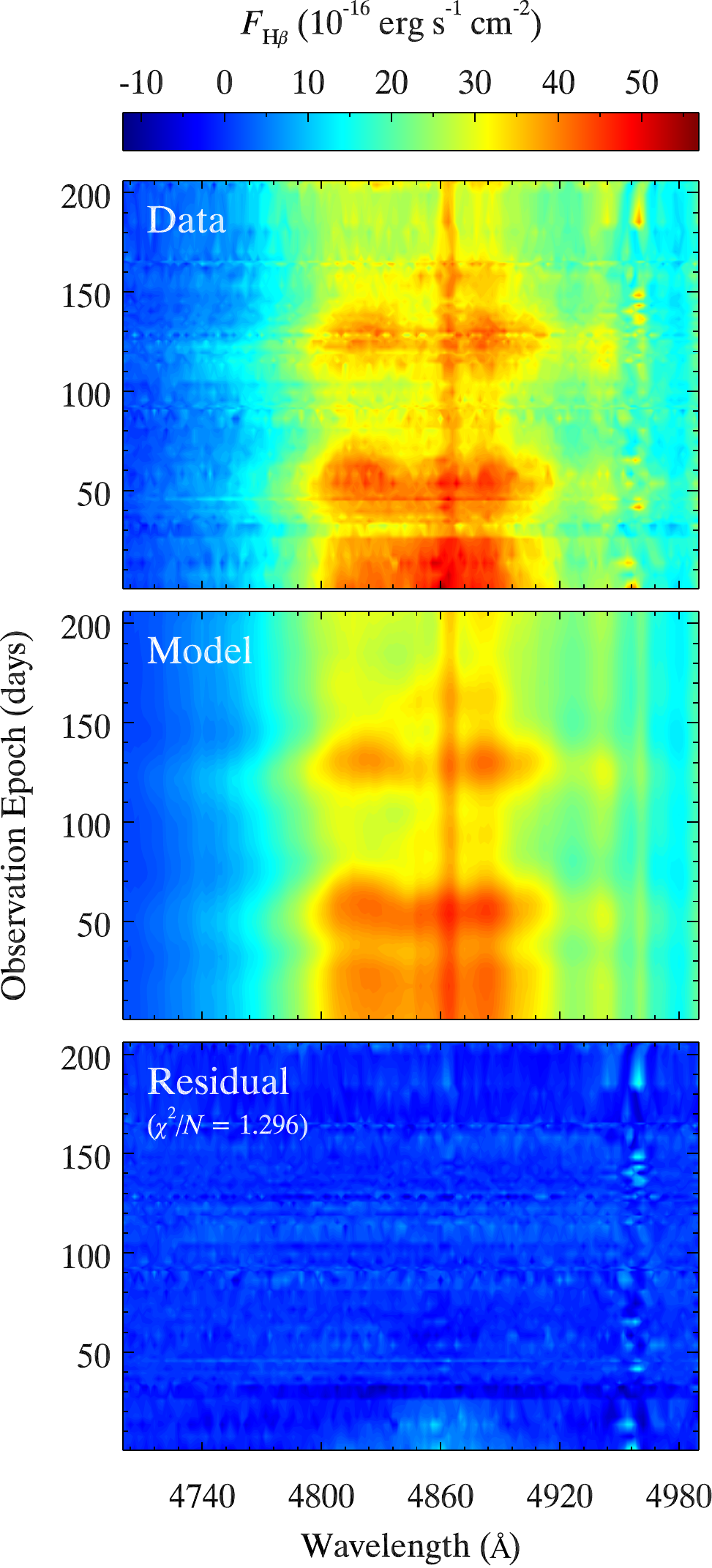} 
	\caption{\footnotesize
		The observed time series of \hb~emission line profiles, the overall 
		MEM fits, and the residual. 
	}
	\label{fit}
\end{figure}
\vglue 0.5cm

\section{MEM fitting}
\label{secanl}
MEM is an effective method for fitting data without assuming any algebraic form of the model.
The method has been applied to the RM observations, and successfully recovered their corresponding 
velocity-delay maps of about 11 sources (see \citealt{ulrich1996,bentz2010b,grier2013,xiao2018}). 
The principle and equations of the MEM have been described in \cite{horne1994} 
and \cite{xiao2018}, and we make a brief overview here. Generally speaking, MEM introduces a 
linearized echo model (Equation \ref{RMmodel}; for convenience, we discretize it)
\begin{equation}
L(v_{i},t_{k})=\bar{L}_0(v_i)+\sum_{j}\varPsi(v_{i},\tau_{j})\left[C(t_{k}
-\tau_{j})-\bar{C}_0\right]\Delta \tau
\label{eq:tf1}
\end{equation}
to fit the observed continuum light curve and the variations of the emission-line profiles. 
The velocity-delay map $\varPsi(v_{i},\tau_{j})$, the continuum $C(t_{k})$, and the 
background spectrum $\bar{L}_0(v_{i})$ are treated as parameterized model in the MEM 
fitting. Here $\bar{L}_0(v_{i})$ is designed to account for the non-variable component of the
emission line, and $\bar{C}_0$ is a reference continuum level, where we adopt the median of the 
continuum flux as $\bar{C}_0$. The MEM fitting is accomplished by minimizing
\begin{equation}
Q={\chi}^{2}-\alpha S,
\label{eq:Q}
\end{equation}
where ${\chi}^{2}=\sum_{m}\left[D_{m}-\mathscr{M}_{m}(\vec{p})\right]^{2} /\sigma_{m}^{2}$ 
constrains the ``goodness-of-fitting'' by ``pulling'' the model $\mathscr{M}_{m}(\vec{p})$ 
toward the data $D_{m}$ ($D_{m}$ and $\mathscr{M}_{m}(\vec{p})$ include the emission line 
and continuum light curves), $S=\sum_{n}\left[p_{n}-q_{n}-p_{n}\ln(p_{n}/q_{n})\right]$ is 
the entropy which controls the ``simplicity-of-modeling'' by minimizing the differences 
between the model parameters $p_{n}$ and the ``default image'' $q_{n}$. Here $m$ and $n$ 
denote the numbers of the observational points and the parameters in the model, respectively.
The parameter $\alpha$ controls a trade-off between the ``goodness-of-fitting" and the 
``simplicity-of-modeling", which means increasing $\alpha$ smooths the MEM model and leads
to larger $\chi^2/N$, and vice versa. The model parameter $p_{n}$ includes $\varPsi(v_{i},\tau_{k})$, 
$C(t_{k})$ and $\bar{L}_0(v_{i})$, and $q_{n}$ is designed as the geometric mean of $p_{n}$.
For one-dimensional model components ($C(t)$ and $\bar{L}_0(v)$), we define
\begin{equation}
q(x)=\sqrt{p(x-\Delta x)p(x+\Delta x)},
%q_{n}=\sqrt{p_{n-1}p_{n+1}},
\end{equation}
where $x$ is $t$ or $v$ for $C(t)$ or $\bar{L}_0(v)$, respectively,
and for two-dimensional model ($\varPsi(v,\tau)$):
\begin{eqnarray}
\label{eq:df2}
\ln{q(v,\tau)} &= &\dfrac{1}{ 1+\calA }\left[\ln{ \sqrt{ p(v-\Delta v,\tau) \, p(v+\Delta v,\tau) } }\right.\nonumber\\
&+&\left. \calA \, \ln{ \sqrt{ p(v,\tau-\Delta\tau) \, p(v,\tau+\Delta\tau) } }\right]
.
\label{q2d}
\end{eqnarray}
Here $\calA$ is a parameter which assigns the weight and controls the aspect ratio of 
$\varPsi(v_{i},\tau_{k})$ in $v$ and $\tau$ direction. 
Increasing $\calA$ smears out the fine structures along the $v$ direction, and vice versa.
In this way, the total entropy can be written as 
$S=(S_{\bar{L}_0}+\calW S_{\varPsi}+\calWW S_{C})/(1+\calW+\calWW)$, where $\calW$ and 
$\calWW$ are weight parameters which control the relative ``stiffness'' of $L(v_{i},t_{k})$, 
$C(t_{k})$ and $\bar{L}_0(v_{i})$. 
In the MEM fitting, $\alpha, \calA, \calW$ and $\calWW$ are the user-controlled parameters. 
The selection of these parameters has been discussed in \cite{xiao2018}.

Similar to \cite{grier2013} and \cite{xiao2018}, we first model the continuum by using the damped 
random walk (DRW, \citealt{li2013, zu2013}), then use the resulting highly sampled continuum in the 
MEM fitting instead of the original one. One benefit of this approach is that the DRW model can give 
reliable uncertainties. Additionally, the DRW model can be used to extrapolate the 
continuum light curve to times shortly before the campaign began, and thus provide 
better constraints on the MEM continuum modeling. 

In Figure \ref{spec}, we demonstrate the MEM fitting of the emission-line light curves at some 
uniformly-spaced (in velocity space) wavelengths, and draw the corresponding one-dimensional transfer 
functions in the left panels. The reconstruction of the continuum is shown in the bottom panel. It is
obvious that the transfer functions of the line wings exhibit simple structures (basically show only 
one dominant peak), whereas the transfer functions around the line core are relatively complex with at 
least two peaks.

In order to better illustrate the overall fitting, we compare the time series of the line 
profiles and the corresponding MEM recovery in Figure \ref{fit}.
In general, the model fits nicely with the \hb~line profiles at all epochs, and the $\chi^2/N$ of the 
overall fitting is 1.296. In addition, the residual (bottom panel of Figure \ref{fit}) shows some weak
signals (at $\sim$ 4959 \AA\ and $\sim$ 4861 \AA), which are coming from the imperfect \oiii\ 
$\lambda$4959 and \hb\ narrow line subtractions.

\begin{figure}
	\centering
	\includegraphics[angle=0,width=0.5\textwidth]{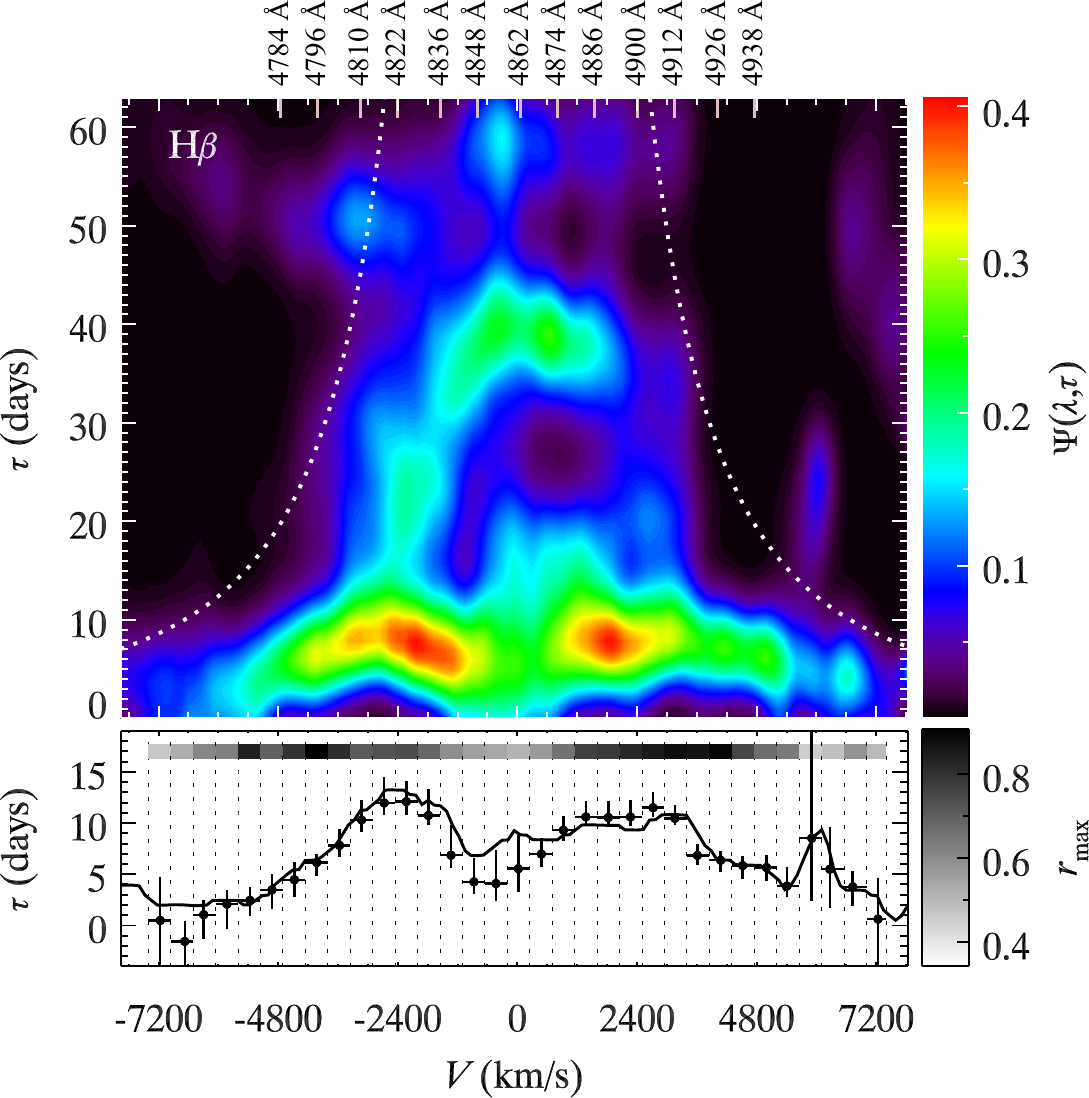} 
	\caption{\footnotesize
		Top panel: \hb~velocity-delay map of NGC~5548. The dotted lines show the ``virial envelope'' 
		$V^2\tau c/G = 8.71\times10^7\sunm$, based on the ``virial product'' measured from the mean 
		time lag and the line dispersion (\citealt{lu2016}). The labels on the top axis are the 
		corresponding wavelengths selected in Figure \ref{spec}. Bottom panel: The centroid time 
		lag of the velocity-delay map (solid line), the velocity-resolved measurements (black dots), 
		and the corresponding maximum cross-correlation coefficients $r_{\rm max}$ (gray-scale bars).
	}
	\label{map}
\end{figure}

\begin{figure*}
	\centering
	\includegraphics[angle=0,width=0.98\textwidth]{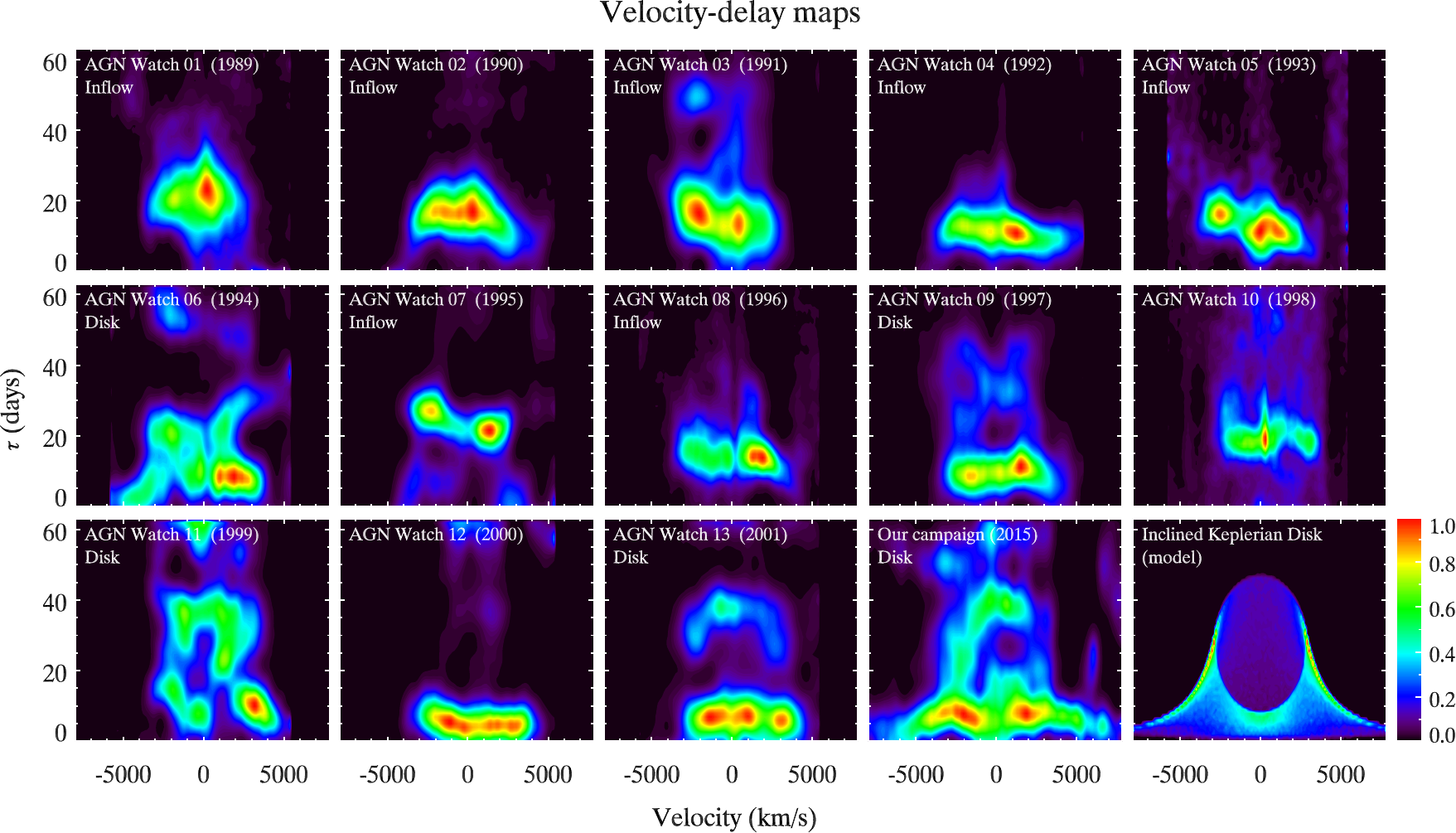} 
	\caption{\footnotesize
		\hb~velocity-delay maps recovered from the AGN Watch data (first 13 panels), the \hb~map 
		of our campaign (penultimate panel), and the simulated map of an inclined Keplerian disk 
		(last panel). We normalize the maps and make their peak values equal to 1. 
	}
	\label{all}
\end{figure*}	

\begin{figure*}
	\centering
	\includegraphics[angle=0,width=0.98\textwidth]{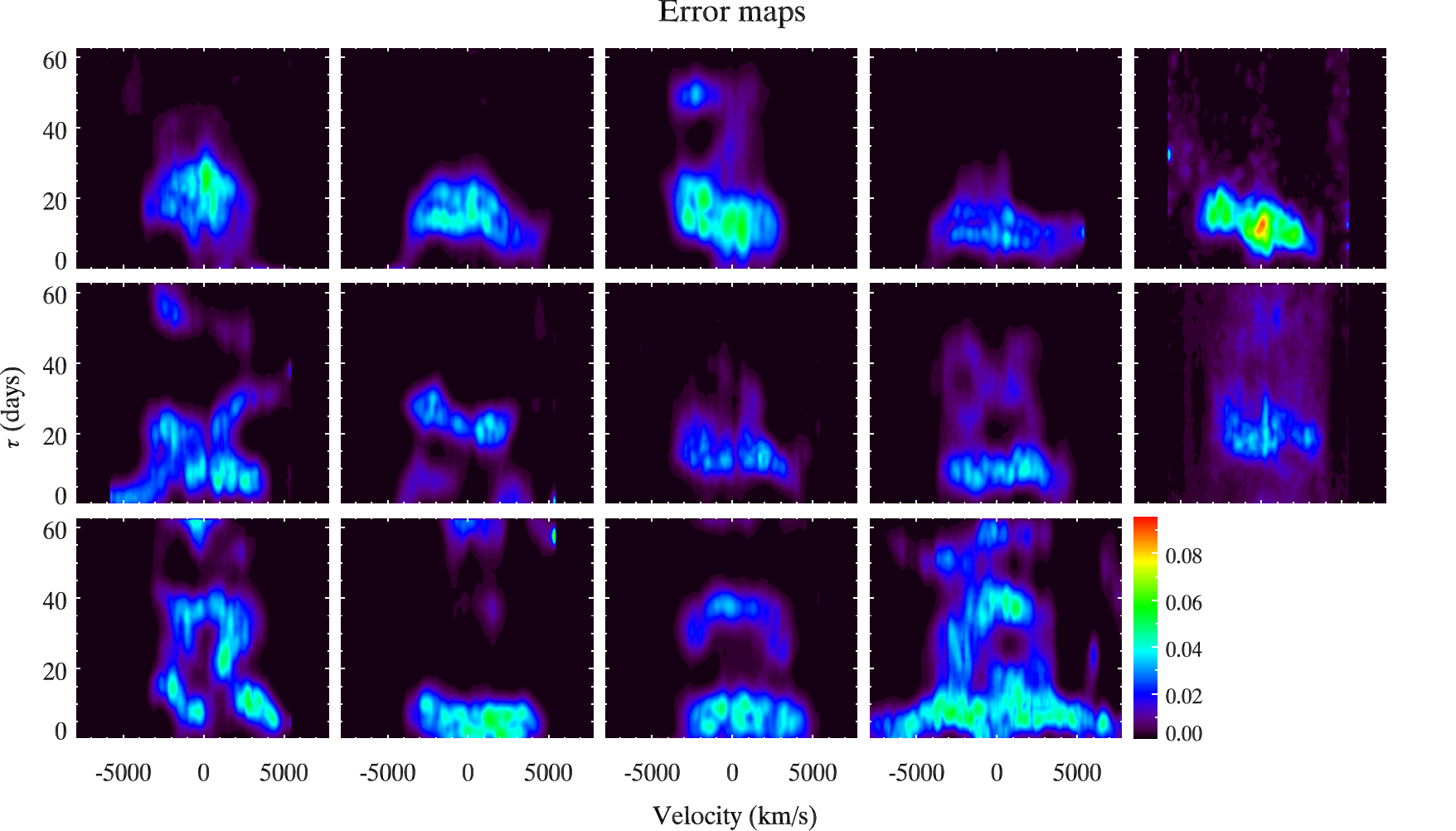} 
	\caption{\footnotesize
		The standard deviations of the velocity-delay maps shown in Figure \ref{all}. 
		%We normalize each panel and make their peak values equal to 1. 
		We normalize the panels by the same factors of Figure \ref{all}.
	}
	\label{fig:std}
\end{figure*}

\section{The Velocity-delay Map}
\label{secvd}
Figure \ref{map} plots the \hb~velocity-delay map of NGC~5548 recovered from our data.
We mark the wavelengths, which are selected to be shown in Figure \ref{spec}, in its top axis.
The map shows a symmetric ``bell'' shape, with a wide velocity distribution 
at short lags ($\lesssim$15~days) and a narrower velocity dispersion at longer lags. The broad wings 
extend to $\sim\pm$7200~$\kms$, while the response in the line core extends to $\sim$ 48 days. In 
addition, there is a hollow in the core of the map, and the response is relatively weaker at 
$\sim [2400~\kms, 28$ days].
For comparison, we plot the ``virial envelope'' $v^2=G\bhm/c\tau$ in Figure \ref{map} with dotted 
lines, where $M_\bullet$ is the black hole mass, $G$ is the gravitational constant, and $c$ is the 
speed of light. Here we adopt $\bhm=8.71\times10^7\sunm$ in \cite{lu2016}. The line response is 
compatible with the envelope, despite the weak ``spine'' at [$v\sim6000\ \kms$, $\tau\sim12-30$\ days] 
and the ``blob'' close to the line core at $\sim$ 60 days which are affected by the residuals coming 
from \oiii~$\lambda4959$ and \hb~narrow line subtractions, respectively. There is a weak response at
$\sim [-3000~\kms, 50$ days], by comparing with the error map in Figure \ref{fig:std} (see 
the next section), this weak feature is significant. Its origin and evolution merit further 
investigations.

In Figure \ref{map}, we also demonstrate a comparison between the mean time lags at different 
velocities in our map and the velocity-resolved time-lag measurements. Here the velocity-resolved 
time lags are derived by dividing the \hb\ profile into 33 uniformly-spaced (450 $\kms$) bins, 
and cross-correlating the \hb\ light curve in each bin with the continuum (see more details in 
\citealt{lu2016}). This result is essentially identical to that presented in \cite{lu2016}, 
although the \hb\ profile is divided into narrower bins. The corresponding maximum 
cross-correlation coefficients ($r_{\rm max}$) are marked by the gray-scale bars. 
To do the comparison, we convolve the velocity-delay map with the autocorrelation function 
(ACF) of the continuum (the output is identical to the CCF), and calculate the centroid around 
the peaks (\textgreater~80\%) of the outputs as the mean time lags. As expected, in general, 
the two results are consistent. The double-peaked structure of the velocity-resolved
lags is similar to what was found by \cite{pei2017} for the 2014 campaign.

Theoretically, a virialized BLR produces a velocity-symmetric signature like ``bell'' shape. 
This is because the velocity-delay structure of a Keplerian orbit is an ellipse, the orbits 
at inner (outer) radii of a virialized BLR produce ellipse structures on the map with wider 
(narrower) velocity distribution at shorter (longer) time delay, and the map is confined 
within the ``virial envelope'' (e.g., \citealt{bentz2010b,grier2013,xiao2018}). 
In particular, the map of an inclined Keplerian disk has a lack of response in the core, 
and is different from the signature of a spherical shell, which has a filled `bell' shape
(e.g., see Figure 1 of \citealt{horne2004} and Figure 14 of \citealt{grier2013}).

The ``bell-like" velocity-delay map of NGC 5548 implies that its BLR is probably an inclined 
disk. It is unlikely to explain the map as a spherical shell geometry, because of the response 
deficit from $\sim$ 20 days to $\sim$ 32 days. The response of the map on the red side at 
$\sim [2400~\kms, 28$ days] is relatively weaker. Considering that the response of this area 
comes from the outer part of the BLR, such evidence indicates that the \hb\ response at the 
outer radius may be more anisotropic or inhomogeneous. 

\section{long-term variability}
\label{secdis}
In order to investigate the long-term variability of the BLR kinematics and geometry, 
we compile the historical spectroscopic data from the AGN Watch archive
\footnote{\url{http://www.astronomy.ohio-state.edu/~agnwatch/}}, which is by far the largest 
optical monitoring project of NGC~5548. This project involves 13 observing campaigns 
from Dec 1988 to Sept 2001, and each campaign has a time span of more than $\sim280$ days. 
The spectra are calibrated by using the \oiii\ narrow emission line as in \cite{peterson2002}.
We did not apply the fitting scheme described in Section \ref{secob} to the AGN Watch data.

In Figure \ref{all}, we present the 13 velocity-delay maps recovered from the AGN Watch data 
together with our map (the same as in Figure \ref{map}). For comparison, we draw a simulated 
velocity-delay map for an inclined Keplerian disk in the bottom right panel of Figure \ref{all}. 
The disk is inclined ($i=45^{\circ}$ to the observer) with an inner radius of $R_{\rm in}=3$ 
lt-days and an outer radius of $R_{\rm out}=28$ lt-days, the emissivity distribution of 
the BLR clouds is assumed to be $\epsilon \propto R^{-1}$. The AGN Watch maps are denoted as 
AGN Watch 01-13. We apply the flux randomization (see details in \citealt{peterson1998}),
which modify the flux of each datum by a random Gaussian deviate within the flux uncertainty,
and use the Monte Carlo (MC) simulations to calculate the uncertainties of the velocity-delay 
maps. The error maps are shown in Figure \ref{fig:std}. Given 
the error maps, the differences between the velocity-delay maps are significant.
From the AGN Watch 01 to 05, the maps show continuing
inflow signatures with longer lags at the blue end and shorter lags toward the red end. 
Interestingly, at this period, the average time lags are generally decreasing as well
(see Table 8 in \citealt{peterson2002}). The AGN Watch 07 to 08 are also dominated by 
inflow signatures with a decrease in the time lags. It implies that the shrink of its 
BLR may correlate to the inflow dynamics. It has been illustrated that the BLR size of 
NGC~5548 follows its continuum luminosity \citep{eser2015}. However, subsequent study 
reveals that the variation of the BLR size lags $2.35^{+3.47}_{-1.25}~\rm yrs$ 
behind the luminosity change, and this lag is similar to the dynamical timescale 
($\sim 2.1~\rm yrs$) of its BLR \citep{lu2016}. The recombination timescale is
insufficient to explain the variation of the BLR time lag, the change of its BLR
kinematics is needed \citep{lu2016}.
The maps of the AGN Watch 06, 09, 11, and 13 generally show symmetric signatures with
a paucity of response in the cores, which are similar to the signature of an inclined 
disk found in our map. They reveal that the BLR is disk-dominated in these periods. 
However, the detections of the response are limited by the data quality.
The rest of the maps are not well recovered due to the low sampling rate 
or the small \hb~variability (see Table 6 and 7 in \citealt{peterson2002}). 
Figure \ref{all} shows a chronological series of the velocity-delay maps, indicating 
transitions between inflows and virialized status. This provides evidence for the BLR 
origin from the tidally disrupted clumps from the torus (\citealt{wang2017}).

\section{summary}
We present the high-quality \hb~velocity-delay map of NGC~5548 recovered from our RM 
campaign in 2015. The map clearly shows a symmetric "bell-like" signature with a lack 
of response in the core. Such a structure is in accord with the predicted map of a 
Keplerian disk. The weaker response in the red than in the blue side at $\sim28$ 
days of the map indicates that the response at the outer radius of the BLR may be 
anisotropic or inhomogeneous. We also show the velocity-delay maps constructed from the 
13-years AGN Watch data. The maps of the seven years reveal that the decreasing BLR size 
is probably related to the inflowing BLR gas. The other four maps show potential disk 
signatures which are consistent with our map.
The velocity-delay maps of NGC 5548 imply that its BLR was switching between the inflow and 
virialized status in the past years. 

\acknowledgements{
We thank the anonymous referee for constructive suggestions.
We thank Hong-Tao Liu for the supports of providing the computing resources, and 
Michael S. Brotherton for his helpful suggestions that improved the manuscript. 
We acknowledge the support of the staff of the Lijiang 2.4m telescope. Funding 
for the telescope has been provided by CAS and the People's Government of Yunnan 
Province. This research is supported by the National Key R\&D Program of China 
(grants 2016YFA0400701 and 2016YFA0400702), by NSFC through grants NSFC-11503026, 
-11233003, -11573026, -11703077, -11773029, by Grant No. QYZDJ-SSWSLH007 from the 
Key Research Program of Frontier Sciences, CAS, and by the Key Research Program 
of the CAS (grant No. KJZD-EW-M06). This work has made use of data from the AGN 
Watch archive.}

%

%\clearpage	
	
\end{document}